# DeepHEN: quantitative prediction essential lncRNA genes and rethinking essentialities of lncRNA genes

Hanlin Zhang[1], Wenzheng Cheng[1]

[1] College of Intelligence and Computing, Tianjin University, Tianjin 300350, China

hanlin_zhang@tju.edu.cn, cwz_1605672099@tju.edu.cn

**Abstract**
Gene essentiality refers to the degree to which a gene is necessary for the survival and reproductive efficacy of a living organism. Although the essentiality of non-coding genes has been documented, there are still aspects of non-coding genes' essentiality that are unknown to us. For example, We do not know the contribution of sequence features and network spatial features to essentiality. As a consequence, in this work, we propose DeepHEN that could answer the above question. By buidling a new lncRNA-proteion-protein network and utilizing both representation learning and graph neural network, we successfully build our DeepHEN models that could predict the essentiality of lncRNA genes. Compared to other methods for predicting the essentiality of lncRNA genes, our DeepHEN model not only tells whether sequence features or network spatial features have a greater influence on essentiality but also addresses the overfitting issue of those methods caused by the low number of essential lncRNA genes, as evidenced by the results of enrichment analysis.

**INTORDUCTION**
Gene essentiality refers to the degree to which a gene is necessary for the survival and reproductive success of a living system. Genes that are indispensable in fulfilling these functions are classified as essential genes[1]. The concept of gene essentiality is dynamic and influenced by the specific context in which it is assessed. It can vary across different genetic backgrounds and environmental conditions. Essential genes play a crucial role in defining the minimal genome of an organism[2] and represent potential targets for the development of anti-cancer or anti-infection drugs[3,4].

Experimental approaches for determining gene essentiality encompass single or multiple-gene knockouts[5], mutagenesis screening[6], and ribonucleic acid (RNA) interference screening[7]. These methods have found widespread application in screening essential genes within single-cell organisms and cell lines derived from complex multicellular organisms[8]. However, due to technical constraints, the majority of reported essential genes are from single-cell organisms like bacteria[9].Thers is a limited documentation of the essentiality of non-coding genes.

Research show that the non-coding regions take up most of the human genome sequences[10]. Hence, there is a need to determine the essentiality of non-coding genes. There are advancements in CRISPR-Cas9 screening[11] techniques that enable the acquisition of genome-wide essentiality information for human cell lines[12]. However, as stated in a recent review, it remains challenging to determine the genome-wide essentiality of all non-coding regions in the human genome[13].

Extensive research has been conducted on computational predictions of essential protein-coding genes. These predictions rely on the properties exhibited by essential protein-coding genes, such as higher expression levels[14], greater conservation compared to non-essential genes[14]. Numerous computational models have been developed in this regard. DeepHE[15], for example, is a deep learning-based method that utilizes both sequence and network features to predict essential protein-coding genes in humans. Additionally, CLEARER[16] is a powerful machine learning-based method that leverages various types of genomic information to identify essential protein-coding genes across different organisms.

However, for non-coding genes, due to the limited number of essential lncRNA[13], it is hard to build computational models. When it comes to lncRNA(long non-coding RNA) genes, there exits two kinds of computational methods. One of them is those that originally target the estimation of coding gene essentialities, which have been extended to include lncRNA genes. For example, The GIC (Gene Importance Calculator) method[17] is a sequence-based scoring scheme for assessing gene essentiality, originally developed for protein-coding genes. The authors of GIC applied their scoring scheme directly to lncRNA genes. While the GIC method has shown success in predicting the essentiality of seven mouse lncRNA genes, it remains challenging to thoroughly evaluate its performance on a broader range of lncRNA genes and across different organisms. Besides, the XGEP[18] method integrated gene expression profiles from numerous cancer cell lines to develop a machine learning-based model for predicting essential protein-coding genes. Similarly to GIC, they also applied the XGEP model directly to lncRNA genes. While there are promising results and supporting evidence in the literature, quantitatively evaluating its performance on lncRNA genes remains challenging. Similar to GIC, the XGEP method also directly applied its model to lncRNA genes. While there are promising results and supporting evidence in the literature, it remains challenging to quantitatively evaluate the performance of the XGEP model on lncRNA genes. The another is those methods that target directly the estimation of non-coding gene essentialities. The SGII method[19], which prioritizes lncRNA genes based on their centralities within the lncRNA-protein-protein interaction network. Using a manually curated dataset of essential lncRNA genes, they observed that essential lncRNA genes tend to exhibit higher betweenness centrality (BC) and degree centrality (DC) compared to the average level of all lncRNA genes. This observation aligns with the characteristics of essential proteins in protein-protein interaction networks. However, a centrality-based scoring scheme is too simplistic to capture the intrinsic structural features of the lncRNA-protein interaction network. Therefore, Zhang *et al*. propose iEssLnc[20] model to solve the problem. iEssLnc is a deep learning model that can quantitative predict the essential lncRNA genes on the genome level. However, suffering from the low number of essential lncRNA genes, the data hungery model, iEssLnc, is a little over-fitting.

LncRNA-protein interactions play a crucial role in determining the molecular functions (MF) of lncRNAs. Consequently, these interactions hold significant potential for predicting lncRNA essentiality. Meanwhile protein–protein interaction is also important in indicating the MF of proteins. As a consequence, we build a lncRNA-protein-protein interaction network.

With the advancement of graph neural network(GNN), GNN has shown its power in representing graph structural information with low-dimensional representations. Hence, GNN has been used in many bioinformatics studies, such as , predicting drug-target interaction[21], predicting lncRNA-protein interaction[22].

Representation learning, also known as feature learning or feature extraction, is a fundamental concept in machine learning and artificial intelligence. It refers to the process of automatically learning meaningful and informative representations or features from raw data. Recently, representation learning demonstrate a greate power in bioinformatics, such as protein learning[23], RNA learning[24]. These power representations have greatly facilitated downstream tasks.

In this work, through taking the advantage of GNN and representation learning, we gain the meaningful representation of both sequence feature and network feature. The dbEssLnc database [73] comprises a comprehensive collection of essential lncRNA genes with literature evidence, encompassing over a hundred genes in both human and mouse. It stands as the largest repository of essential lncRNA genes to date. Based on the idea of self-supervised clustering and dbEssLnc database, we build a benchmarking dataset to train our supervised classification model. The result of enrichment analysis indicate that our model has a better performance over other methods. We also analysis the contribution of sequence features and network spatial features to essentiality.

## MATERIALS AND METHODS

### The workflow of DeepHEN model

The overview of DeepHEN is shown in Figure 1. In this research, we first constructed a lncRNA-protein-protein interaction (LPPI) network. We used graph embedding methods to capture network feature of lncRNA genes. We also utilized dna2vec[25] to capture sequence feature. Due to the limited amount essential lncRNAs, we conducted negative sample optimization depending on the idea of semi-supervised clustering. Then supervised machine-learning model was trained. Many essential lncRNAs were predicted by our DeepHEN model. The correctness of DeepHEN model was verified by enrichment analysis method.

### Dataset Curation

The human protein-protein interaction data from BioGrid database version 4.4[26] was downloaded. The PPI network includes 32142 proteins and 664819 protein-protein interactions. The human lncRNA-protein interactions were obtained from the NPI database v4.0[27] to built lncRNA-protein interaction (LPI) network. By cross-referencing the names of proteins in both datasets, the lncRNA-protein interactions and protein-protein interactions were integrated, resulting in a heterogeneous network comprising two types of interactions. This network was named the LPPI network. In order to focus on lncRNAs, the proteins that do not interact with lncRNAs were removed. After processing, the LPPI network contains a total of 41530 lncRNAs, 3003 proteins, and 485087 interactions. To obtain the lncRNA gene sequence needed for our research, we finally acquired 29482 human lncRNAs.

The PPI network can be defined as $P = (U, I)$, in which $U$ is the set of protein nodes, and $I$ defines edges that represent the protein-protein interactions. Besides, the LPPI network can be formulated as a heterogeneous graph $G = (V, E, T)$, in which each node $v \in V$, and each interaction $e \in E$, are associated with their mapping functions $\varphi(v): V \rightarrow T_V$, and $\phi(e): E \rightarrow T_E$ respectively. The set $T_V \subseteq T$ and the set $T_E \subseteq T$ contain node types and edge types in $G$. In LPPI network, the set $T_V$ = {lncRNA, protein}, and the set $T_E$ = {lncRNA-protein, protein-protein}. Therefore, we have:

$$|T_V| + |T_E| > 2 \qquad (1)$$

Where |.| is the cardinal operator in the set theory. This makes $G$ a heterogenous graph. We gained lncRNA gene sequences from SGII study and 154 human essential lncRNA genes from the dbEssLnc database[28].

### Sequence features learning

The traditional way to extract features from lncRNA gene sequence is to calculate features such as, Gene Importance Calculator (GIC) or sequence length that mainly depends on human knowledge. However, due to the lack of human knowledge, traditional way can not capture enough features from lncRNA gene sequence for downstream task. Therefore, we used dna2vec model to solve this problem. Like network embedding mentioned above, the dna2vec model can also learn continuous low-dimensional feature representations of k-mers that are expected to capture the similarity between k-mers.

The first step of dna2vec is to separate genome into long non-overlapping DNA fragments. Next, dna2vec model converts long DNA fragments into overlapping variable-length k-mers that can be considered as corpus. Dna2vec model puts this corpus into the skip-gram of word2vec[29], gaining embeddings of k-mers, which maximizes the probability of retaining neighboring k-mers. In this research, we used the pretrained dna2vec results by hg38 dataset.

Using dna2vec can only get embeddings of k-mers, however, we need to extract feature from sequence. Therefore, we first convert human lncRNA sequences into overlapping 3-mers. For every lncRNA gene sequence, we have $\mathbf{X} = (x_1, x_2, x_3\ldots\ldots x_n) \in \mathbb{R}^{n \times d}$, where $x_i$ is the dna2vec pretrained result of $i$-$th$ 3-mers, $n$ is the number of non-overlapped 3-mers of lncRNA gene sequence, d is the dimension of dna2vec pretrained result and d is 100. Only summing the embeddings of every 3-mers of lncRNA gene sequence together according to the pretrained dna2vec result will omit positional information. Therefore, inspired by the great transformer[30], we utilize positional encoding in our sequence learning part. The positional matrix $\mathbf{P}$ is defined as follows:

$$P_{i,2j} = 0.01 sin(\frac{i}{10000^{2j/d}}) \quad (2)$$

$$P_{i,2j+1} = 0.01 cos(\frac{i}{10000^{2j/d}}) \quad (3)$$

Where $i$ is the $i$-$th$ 3-mers of lncRNA gene sequence and $2j$, $2j+1$ are the $2j$-$th$, $2j+1$-$th$ dimensions of the dna2vec pretrained results. The lncRNA gene sequence feature matrix $\mathbf{Y}$ is defined as follows:

$$\mathbf{Y} = \mathbf{X} + \mathbf{P} \quad (4)$$

Then the lncRNA gene sequence feature is calculated by adding up each row of the matrix $\mathbf{Y}$.

Sequence features are also extracted from proteins in our research. To learn protein sequence features, we utilize the method of lncRNA gene feature learning and employ ProtVec[31]. ProtVec generates three lists of shifted non-overlapping words for each protein sequence. These lists are then split into non-overlapping fixed-length 3-mers. The generated corpus is finally fed into the skip-gram of node2vec to learn the representation of protein 3-mers. In this research, we use the pretrained ProtVec results using the Swiss-Prot dataset.

In this research, each protein sequence is converted into overlapping 3-mers. Next, we sum the embeddings of every 3-mers of protein sequence together according to the pretrained ProtVec results. The dimension of the protein sequence feature is controlled to 100, which is consistent with the lncRNA gene sequence feature.

## Network features learning

Topology features typically only extract one type of network topology feature and cannot capture the similarity between nodes in the network. To overcome this limitation, we used the Variational Graph Auto-Encoders (VGAE)[32] to learn embeddings of nodes in the LPPI network and obtain continuous low-dimensional feature representations of each node.

We first assigned features to nodes in the LPPI network. For nodes $v \in V$ and $\varphi(v)$ = protein, we used ProtVec to assign features as mentioned above. For nodes $v \in V$ and $\varphi(v)$ = lncRNA, we initialized the features of these nodes with a normal distribution. We wrote the resulting graph of feature assignments into the node feature matrix X and the adjacency matrix A, which serve as the input for subsequent modeling. Next, we used the VGAE model to learn a latent, interpretable representation of the input matrix data. VGAE consists of two parts: an inference model and a generation model.

In the inference model, VGAE learns to generate corresponding latent variables Z for each input variable by learning a normal distribution for each input variable. Z is represented as a Gaussian distribution, as shown below:

$$q(Z|X, A) = \prod_{i=1}^{N} q(z_i | X, A) = \mathcal{N}(z_i | \mu_i, \text{diag}(\sigma_i^2)) \tag{5}$$

Where $q(Z|X, A)$ refers to the encoder function that maps each input variable to its corresponding mean and variance vectors, denoted as $\mu_i$ and $\sigma_i$ respectively. N is the number of nodes in the LPPI network. And the hidden variable Z is sampled from a multivariate Gaussian distribution, where the mean (denoted as $\mu$) and variance (denoted as $\sigma$) are derived from two 2-layer Graph Convolutional Network (GCN) models. These two GCN layers are defined as follows:

$$GCN(X, A) = \tilde{A} \, \text{ReLU}(\tilde{A} X W_0) W_1 \tag{6}$$

Where X is the node feature matrix, $W_0$ and $W_1$ are weight matrices updated by GCN learning.

The ReLU activation function and symmetrically normalized adjacency matrix $\tilde{A}$ are defined as follows:

$$\text{RuLE}(x) = \max(0, x) \tag{7}$$

$$\tilde{A} = D^{-\frac{1}{2}} A D^{-\frac{1}{2}} \tag{8}$$

Where D is the degree matrix. We can then obtain the matrix of mean node vector representations, and similarly, the matrix of variance node vector representations. The GCN layer for calculating the mean vector and the GCN layer for calculating the variance vector share the same $W_0$.

In the generative model section, the reconstructed graph is calculated by taking the inner product between latent variables, as shown below:

$$p(A|Z) = \prod_{i=1}^{N} \prod_{j=1}^{N} p(a_{ij} | z_i, z_j) \tag{9}$$

$$p(a_{ij} = 1 | z_i, z_j) = \sigma(z_i^T, z_j) \tag{10}$$

Where $p(A|Z)$ is the decoder function which transformed the hidden variable Z sampled from the encoder function back to the adjacent matrix A, where $a_{ij}$ are the elements of A, and $\sigma$ is the logistic sigmoid activate function, defined as follows:

$$\sigma(x) = \frac{1}{1+e^{-x}} \tag{11}$$

Similar to VAE, the loss function of VGAE is defined as follows:

$$L = \mathbb{E}_{q(Z|X,A)}[\log p(A|Z)] - KL[q(Z|X,A) \| p(Z)] \tag{12}$$

Where $KL[q(\cdot)\|p(\cdot)]$ is the Kullback-Leibler divergence between $q(\cdot)$ and $p(\cdot)$, and $p(Z)$ is the Gaussian prior, defined as follows:

$$p(Z) = \prod_{i=1}^{N} p(z_i) = \prod_{i=1}^{N} \mathcal{N}(z_i | 0, I) \tag{13}$$

The first term of the loss function L aims to minimize the difference between the reconstructed graph and the original graph. Meanwhile, the second term of the loss function L aims to make the learned distribution of latent variables similar to the gaussian prior. The KL divergence of Equation (12) restricts the posterior distribution to be close to the prior distribution we set beforehand, which acts as a regularization term to prohibit overfitting.

Overall, the inference model encodes real samples as low-dimensional vector representations (latent variables) and learns the distribution of latent variables. The generation model samples the corresponding latent variables of real samples from the distribution of latent variables and generates data that closely matches the real samples. Gaussian noise is introduced in this process to ensure the model's generative capability, and the reparameterization trick is used to replace the sampling process and ensure that the model is trainable. Finally, when the model is trained, we obtain the feature embeddings of LPPI network nodes through the inference model.

**Build benchmarking dataset**

To construct a high-quality benchmark dataset, we extracted only 154 necessary lncRNA genes from the dbEssLnc database and marked the remaining lncRNA genes as unknown. To further filter out non-essential genes and annotate the importance of each gene, we connected feature vectors obtained from both the LPPI network and dna2vec to form a total feature vector for each lncRNA gene.

For each unknown lncRNA gene, we calculated the cosine similarity between its total feature vector and the total feature vectors of each necessary lncRNA gene. We then computed the average of all cosine similarity results to obtain the importance annotation. Based on their importance annotations, we ranked the unknown lncRNA genes and selected the lowest ranked ones to determine the most suitable non-essential lncRNA group. This method of selecting non-essential lncRNAs is known as the negative sample selection method.

Finally, we selected the top 154 best non-essential lncRNA genes to construct the benchmark dataset, and we will describe in detail in the later discussion the validation method used to construct the benchmark dataset.

**LncRNA gene essentiality estimations**

In the supervised machine-learning part of the DeepHEN model, we trained a Support Vector Machine (SVM) to predict the essentiality of lncRNA genes. Our SVM has both a binary classification mode and a quantitative scoring mode. The binary classification mode outputs a binary decision of whether a lncRNA gene is essential or non-essential. The quantitative scoring mode outputs an essentiality score, which is the decision function before applying the final cutoff. The score can be interpreted as the estimated likelihood of a lncRNA gene being essential, before applying the final cutoff.

**Essential lncRNA gene enrichment score**

We calculated the essential lncRNA gene enrichment score using the Kolmogorov-Smirnov-like statistics. First, we sorted all lncRNA genes according to their quantitative essentiality score in descending order. For the $k$-th lncRNA gene in the sorted list, the statistic $l_k$ ($k = 1, 2, …, N$) is defined as follows:

$$l_k = l_{k-1} + \frac{1}{N_e} i_k - \frac{1}{N - N_e}(1 - i_k) \quad (14)$$

Where $N_e$ is the number of essential lncRNA genes, $N$ is the total number of lncRNA genes in the sorted list, and $i_k$ is equal to 1 if the $k$-th lncRNA gene in the list is essential, otherwise $i_k$ is equal to 0. The value of $l_0$ is 0, since it represents the enrichment score for the empty set of genes. The enrichment score ($ES$) is defined as the maximal value of $l_k$, which reflects the degree of enrichment of essential lncRNA genes in the top-ranked genes, as shown below:

$$ES = \max_{1 \leq k \leq N} l_k \quad (15)$$

A larger ES indicates a better performance of the DeepHEN model in predicting essential lncRNA genes. The ES ranges from 0 to 1, with 1 indicating that all essential lncRNA genes are ranked at the top of the list, and 0 indicating that the list contains no essential lncRNA genes.

**Evaluation metrics**

In this research, we used precision(pre), false positive rate (FPR), recall(sensitivity), accuracy and the Matthews correction coefficient (MCC) to evaluate the performance of our classifier. These metrics are defined as follows:

$$\Pr ecision = \frac{TP}{TP + FP} \quad (16)$$

$$FPR = \frac{FP}{TN + FP} \quad (17)$$

$$Sensitivity = \frac{TP}{TP + FN} \quad (18)$$

$$Accuracy = \frac{TP + TN}{TP + TN + FP + FN} \quad (19)$$

$$F1-score = \frac{2*pre*recall}{pre+recall} \qquad (20)$$

Where TP, TN, FP ,and FN are the number of true positives, true negatives, false positives, and false negatives respectively.

In addition, we used the area under the receiver operating characteristic curve (AUROC) to evaluate the overall performance of our classifier. An ROC curve plots the true positive rate (sensitivity) against the FPR for all possible classification thresholds. The AUROC assesses the classifier's ability to distinguish between positive and negative samples, with a higher score indicating better performance.

We also used the area under the precision-recall curve (AUPR) to evaluate the performance of our classifier. A PR curve plots the precision against the recall for all possible classification thresholds. The AUPR evaluates classifiers, with a higher AUPR indicates better performance in terms of the trade-off between precision and recall.

Together, these metrics provide a comprehensive and robust evaluation of the performance of our DeepHEN classifier in predicting essential lncRNA genes.

**Parameter calibrations**

The parameters used in our research were adjusted as follows. We used the PyG package in Python to implement VGAE. The hidden size of the GCN layer in VGAE was set to 128. The dimension of the latent variable was set to 64. The learning rate was set to 0.01, and the training epoch was set to 700. We implemented the SVM predictors using the LIBSVM[33] library in Python. The kernel type of our SVM was set to radial basis function. The gamma value was set to 0.07, and the C value was set to 0.0585.

**Results and discussions**

**Prediction performance analysis and comparisons**

We estimated the prediction performance of the DeepHEN model using 5-fold cross-validations. Our benchmarking dataset introduced information leaks to the training procedures of SGII, GIC, and XGPE methods, against which we compared the performance of our DeepHEN model. Despite this disadvantage, our DeepHEN model outperformed the other methods on all evaluation metrics. The XGEP techniques, which were trained using crucial protein-coding genes, demonstrated strong performance in identifying non-essential lncRNA genes with high precision and low false positive rates (FPR). However, their performance experienced a significant decline when recognizing essential lncRNA genes, with substantially lower values in terms of sensitivity, F1-score, and accuracy. This decline could be attributed to the limited knowledge available on essential lncRNA genes at the time the methods were developed, or to potential inherent variations in essentiality between coding and non-coding genes, a phenomenon that remains unexplored within the life sciences community. Figure 2 shows the performance of DeepHEN in predicting values, as well as comparisons to the SGII, GIC, and XGEP techniques.

To provide a comprehensive evaluation of the prediction performance of DeepHEN, we presented the ROC and PR curves in Figure 3 and computed the corresponding AUROC and AUPR values. The results of cross-validation, ROC curves, and PR curves demonstrate strong performance. However, our benchmarking dataset is small, and during its composition, we selected a portion of unannotated lncRNA genes as negative samples using cosine similarity. This selection process may introduce bias and produce misleading results in cross-validation. Therefore, it is important to verify the usefulness of DeepHEN and the validity of our negative sample selection approach.

**Enrichment analysis and Validation of non-essential lncRNA gene sample strategy**

Our goal is to classify essential human lncRNA, but there is currently no clear definition of human non-essential lncRNA. To compose our benchmarking dataset, we opted for a negative sample selection approach based on the idea of semi-supervised clustering. However, this approach lacks a clear biological basis, and thus its validity needs to be verified. Additionally, while our benchmarking dataset is balanced, in the real world, the number of non-essential lncRNA is likely to be higher than the number of essential lncRNA. Moreover, our benchmarking dataset is relatively small, which may lead to a high number of false positives when our DeepHEN model is applied at the genome level. Therefore, it is crucial to verify the validity of our negative sample selection approach and to test the ability of our DeepHEN model to accurately predict the essentiality of lncRNA at the genome level.

As mentioned earlier, the DeepHEN model outputs an essentiality score for every lncRNA gene. To evaluate the performance of DeepHEN in quantitative scoring mode, we conducted an enrichment analysis of all essential lncRNA genes. We calculated the enrichment score for each lncRNA gene using equation (15) at the genome level and compared it to the enrichment scores calculated for SGII and GIC methods. Figure 4 shows the enrichment curve for DeepHEN, SGII, and GIC methods. In addition to the enrichment curve, we also considered the essentiality score, so our DeepHEN models enable quantitative estimation of the essentiality of lncRNA genes in genome-wide screenings. Our results demonstrate that the essentiality score of DeepHEN is higher than that of other methods, and that essential lncRNA genes tend to have higher ranks in the DeepHEN model. It is worth noting that in the enrichment analysis of the iEssLnc method, these essential non-coding genes tend to be predominantly distributed towards the front portion of the sort, which is too good to be ture. However, our DeepHEN models address the issue.

To further validate our negative sample strategy for non-essential lncRNA genes, we conducted additional experiments. As mentioned earlier, we sorted lncRNAs with unknown essentiality status in descending order of rank and selected the top-ranked lncRNAs as non-essential positive samples. We trained an SVM using this positive sample approach and evaluated its prediction performance, as shown in Table 1. We also conducted an enrichment analysis, as shown in Figure 4. Our results demonstrate that the prediction performance decreases with the positive sample approach, as evidenced by the lower enrichment score and less distinct concentration of essential genes at the genomic level. These findings indicate that the positive sample approach can make the trained predictor unreliable. Therefore, we can confidently conclude that our negative sample strategy for non-essential lncRNA genes is reasonable.

**Sequence and Network Analysis**

Although our DeepHEN model utilizes both sequence and network features of lncRNA genes, it is important to further explore the characteristics of these genes by studying the correlation between lncRNA essentiality and each feature type separately. To this end, we conducted two additional pipelines: Sequence_DeepHEN, which only uses sequence features, and Network_DeepHEN, which only uses network features. To ensure the accuracy of our analysis, we used the same SVM parameters for Sequence_DeepHEN and Network_DeepHEN as for DeepHEN, using the negative sample approach. We evaluated the prediction performance of Sequence_DeepHEN and Network_DeepHEN using both negative and positive sample approaches and plotted the corresponding ROC and PR curves, as shown in Figure 3. We also conducted an enrichment analysis for both approaches using both negative and positive sample approaches, as shown in Figure 5, and present the prediction performance results in Table 1. As the result shows, both sequence feature and network feature contribute to the essentiality. To futher analysis, we find network feature contribute more to the essentialities of lncRNA genes. However, when only considering network feature, our model can not have an impressive proformance.

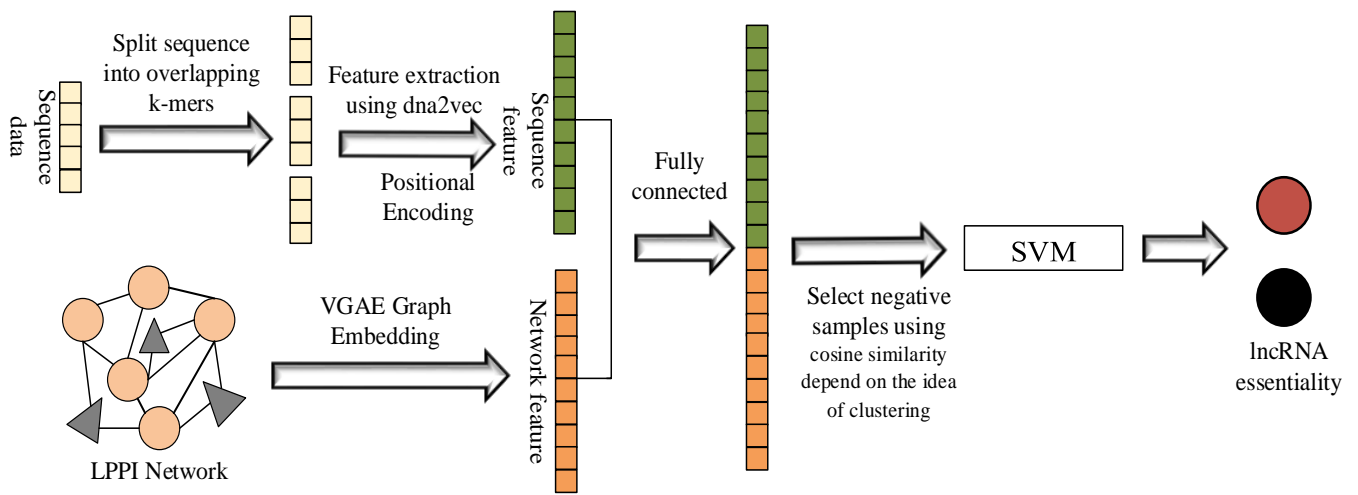

Figure 1 | The workflow of DeepHEN model.

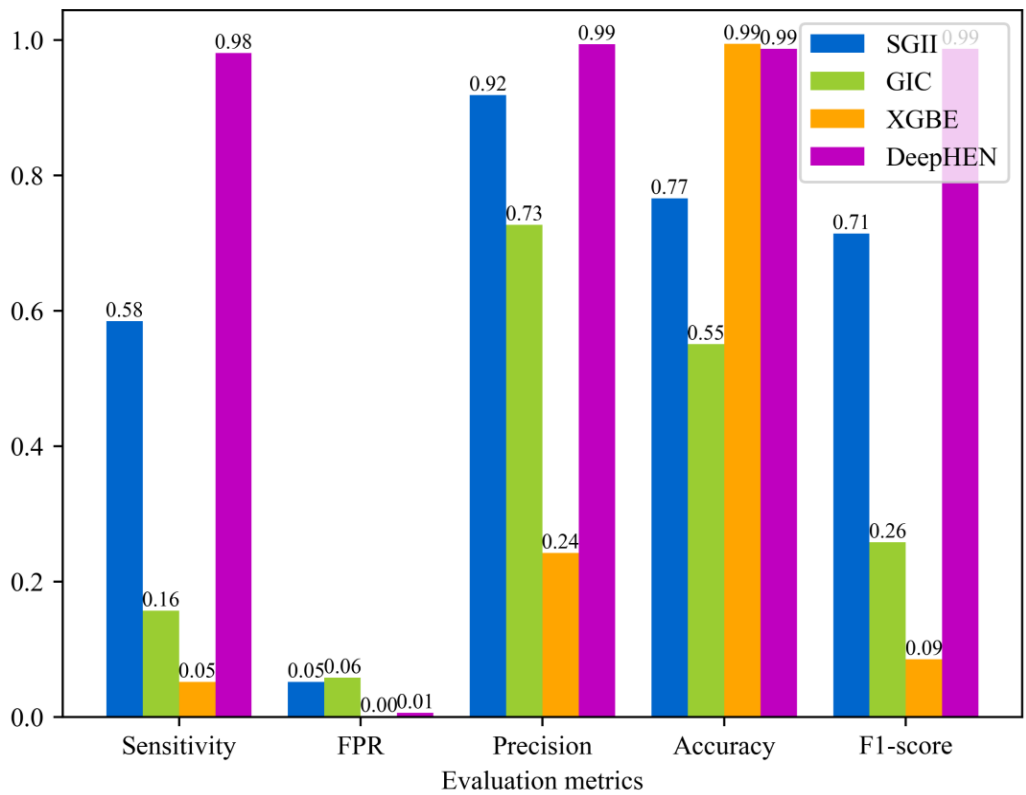

Figure 2 | The prediction performance comparison of DeepHEN, SGII, GIC, XGEP methods.

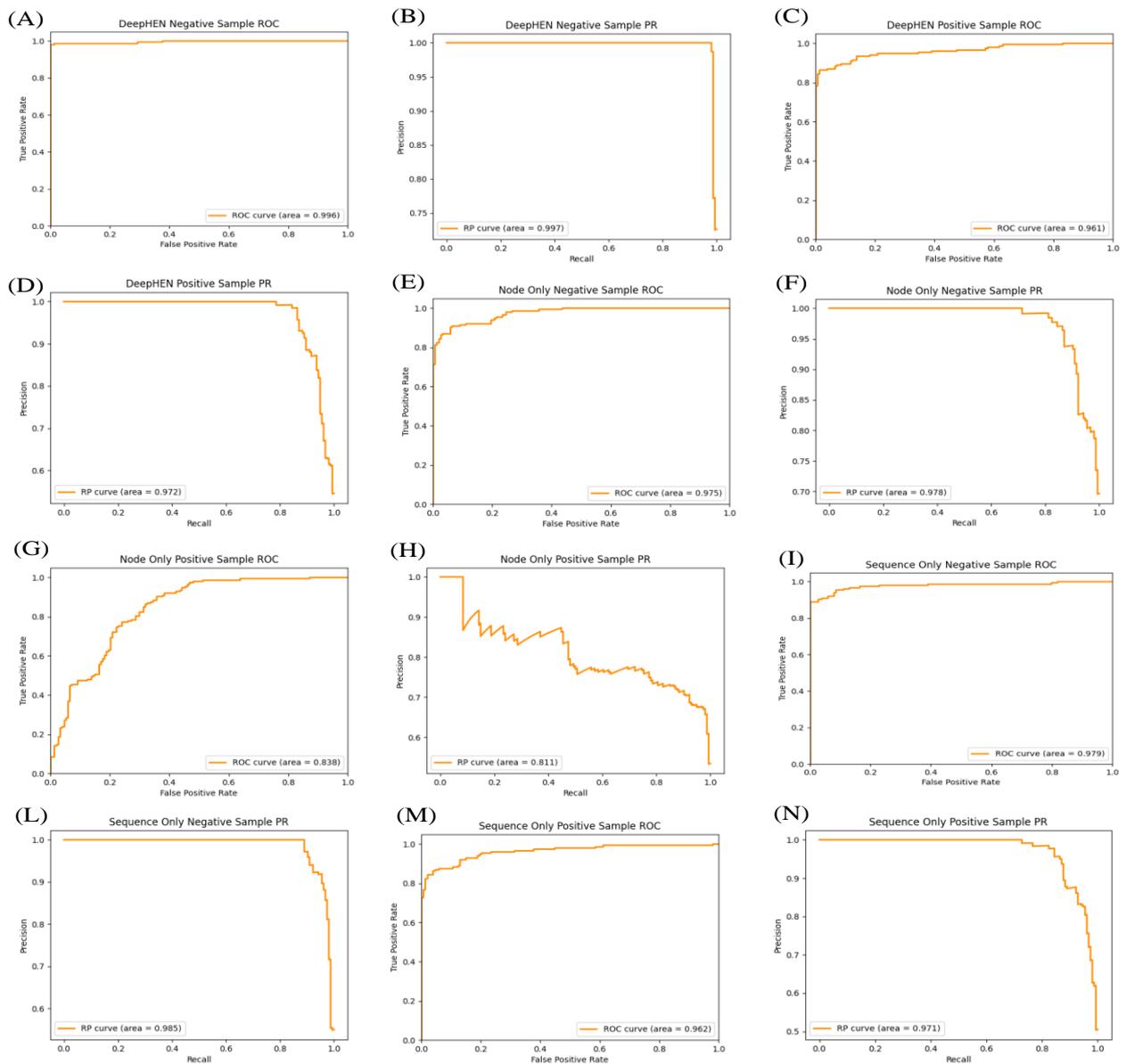

Figure 3 | (A) and (C) show the Receiver Operating Characteristic (ROC) curves of DeepHEN, using negative and positive samples respectively. (B) and (D) show the Precision-Recall curves of DeepHEN, again using negative and positive samples respectively. Similarly, (E) and (G) show the ROC curves of Network_DeepHEN using negative and positive samples, while (F) and (H) show the Precision-Recall curves. Finally, (I) and (M) show the ROC curves of Sequence_DeepHEN using negative and positive samples, and (L) and (N) show the Precision-Recall curves.

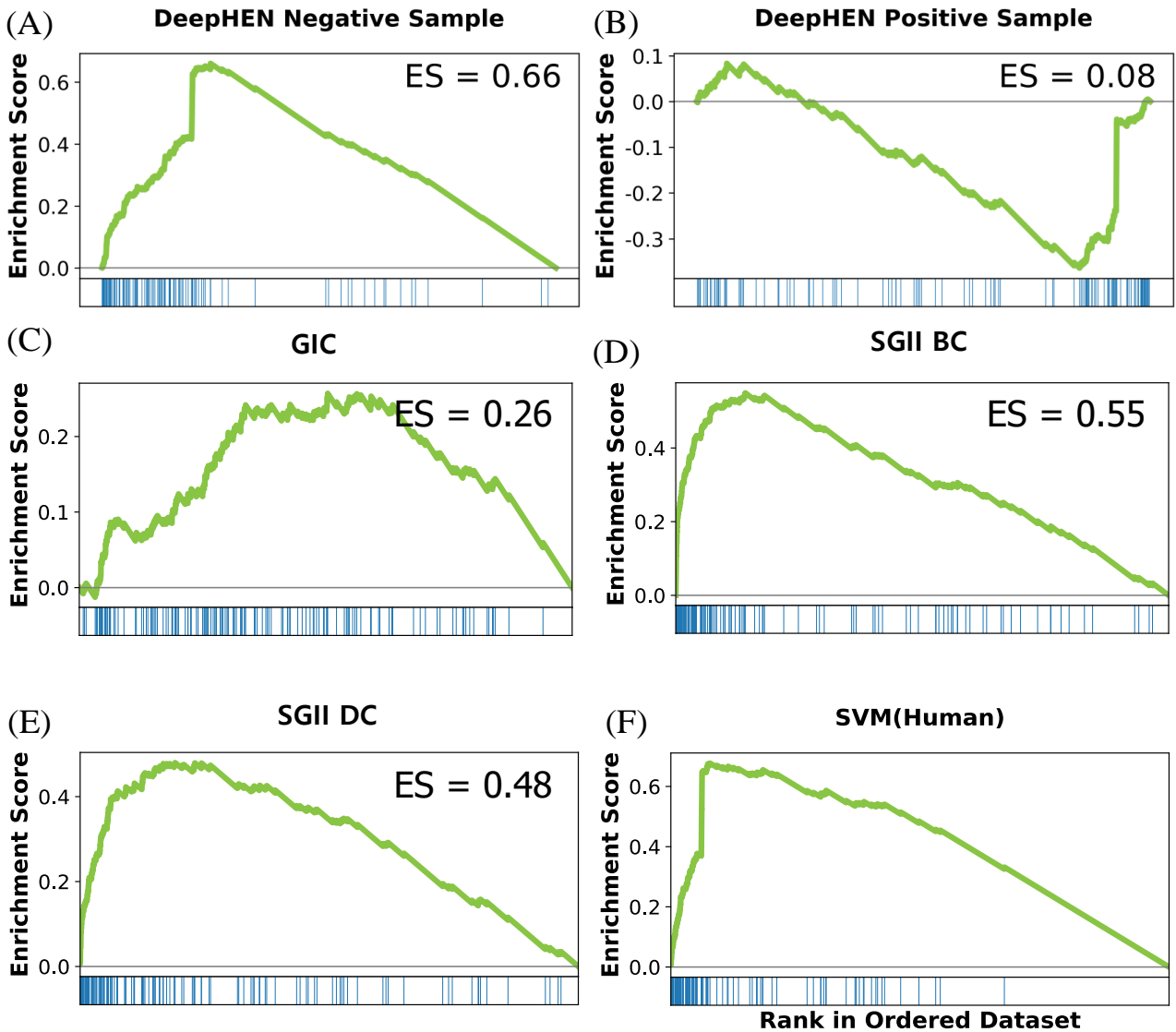

Figure 4 | (A) and (B) show the results of enrichment analysis of DeepHEN in a quantitative scoring mode, using both negative samples and positive samples. The curves represent the running statistics of the enrichment score, plotted against the rank of the sorted list according to the corresponding score. The maximum value of the curve is the final value of the enrichment score, which is indicated in the top-right corner of each panel. (C) shows the results of enrichment analysis of GIC.(D) and (E) show the results of an enrichment analysis of SGII using BC and DC score. (F) show the results of enrichment analysis of iEssLnc.

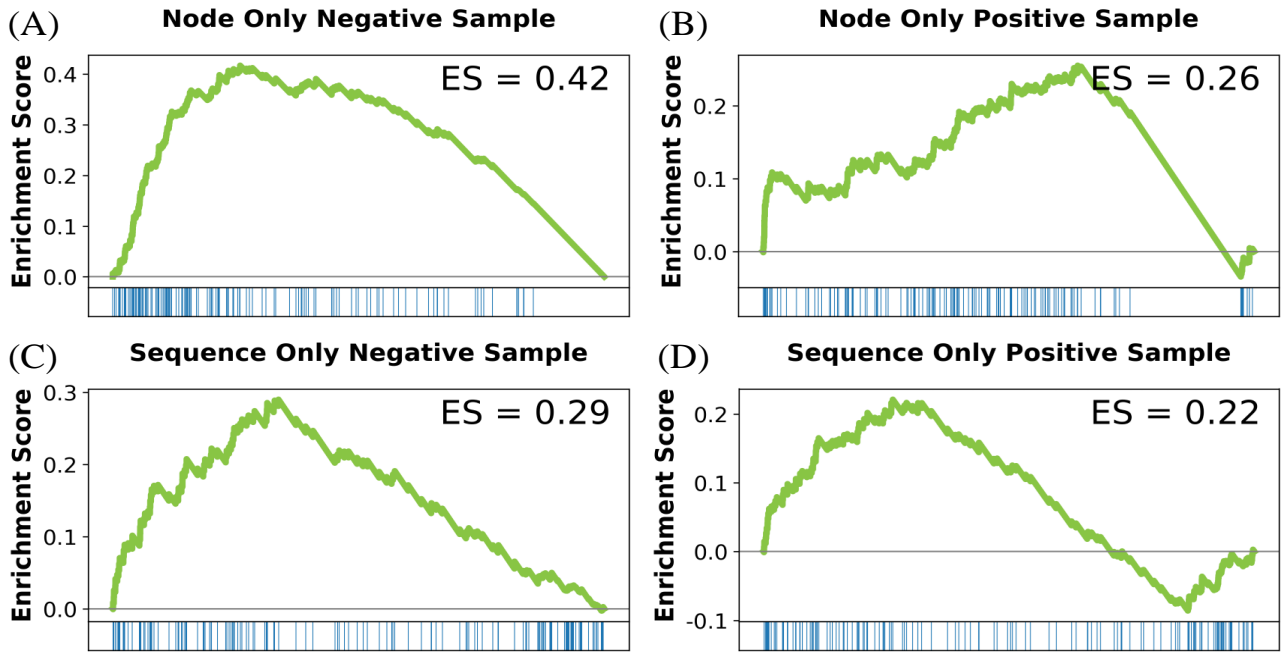

Figure 5 |(A) and (C) show the results of an enrichment analysis of Network_DeepHEN and Sequence_DeepHEN, respectively, using a quantitative scoring mode and negative samples.(B) and (D) show the results of an enrichment analysis of Sequence_DeepHEN using a quantitative scoring mode and positive samples.

Table 1 | The performance of DeepHEN, Sequence_DeepHEN, Network_DeepHEN using positive sample and negative sample.

| Feature | Sample | Sensitivity | FPR | Precision | Accuracy | F1-score |
|---|---|---|---|---|---|---|
| Network&Sequence | Positive | 0.870 | 0.065 | 0.931 | 0.903 | 0.899 |
|  | Negative | 0.981 | 0.006 | 0.993 | 0.987 | 0.987 |
| Network Only | Positive | 0.877 | 0.338 | 0.722 | 0.769 | 0.792 |
|  | Negative | 0.863 | 0.026 | 0.971 | 0.919 | 0.914 |
| Sequence Only | Positive | 0.903 | 0.130 | 0.874 | 0.886 | 0.888 |
|  | Negative | 0.831 | 0.000 | 1.000 | 0.916 | 0.908 |